\documentstyle[12pt,epsfig]{article}
\begin{document}

%%%%%%%%%%%%%%%%%%%%%%%%%%%%%%%%%%%%%%%%%%%%%%%%%%%%%%%%%%%%%%%%%%%%%%%%

% HEADER starts here

% We are supposed to thank the government first...
\renewcommand{\thefootnote}{\fnsymbol{footnote}}
\setcounter{footnote}{0}

\begin{flushright}
IKDA 96/6 \\
hep-ph/9602397 \\
February 1996
\end{flushright}

\begin{center}

{\Large\bf
A Monte Carlo algorithm for Multiphoton\\[1ex]
Beamstrahlung in Monte Carlo event generators%
\footnote{Contribution to the Proceedings of the Workshop on
{\it Physics with $e^+e^-$ Linear Colliders,}
Annecy, Gran Sasso, Hamburg, February 4 to September 1, 1995.}%
\footnote{Supported by Bundesministerium f\"ur Bildung, Wissenschaft,
          Forschung und Technologie, Germany.}
}

\vspace*{0.3cm}

%%% Author:
{\large Harald Anlauf} \\[1ex]
% short:
% {\it TH Darmstadt, Darmstadt, Germany, and
% Universit\"at Siegen, Siegen, Germany}
% longer:
{\it
Technische Hochschule Darmstadt, 64289 Darmstadt, Germany \\[1ex]
and \\[1ex]
Universit\"at Siegen, 57076 Siegen, Germany
}
\end{center}

%%%

\renewcommand{\thefootnote}{\arabic{footnote}}
\setcounter{footnote}{0}

\baselineskip=18pt

%%%%%%%%%%%%%%%%%%%%%%%%%%%%%%%%%%%%%%%%%%%%%%%%%%%%%%%%%%%%%%%%%%%%%%%%

% TEXT BEGINS HERE

\begin{abstract}
We describe a simple algorithm that calculates the distributions of
electrons and positrons under multiphoton beamstrahlung at a future
linear collider.  The evolution equation as given by Chen is solved by a
Monte Carlo algorithm.  Explicit multiple beamstrahlung photons are
generated.  We present first results from an implementation of
beamstrahlung into the Monte Carlo event generator {\tt WOPPER}, that
calculates the QED radiative corrections to the process $e^+e^- \to
4$~fermions through resonating W pairs.
\end{abstract}

%%%%%%%%%%%%%%%%%%%%%%%%%%%%%%%%%%%%%%%%%%%%%%%%%%%%%%%%%%%%%%%%%%%%%%%%

\section{Introduction}

At future linear $e^+e^-$ colliders with center of mass energies of 0.5
TeV and above, beamstrahlung \cite{overview,Chen88}, the synchrotron
radiation from the colliding $e^+e^-$ beams, will become of significant
physical interest.  On one hand, beamstrahlung may carry away a
substantial fraction of primary beam energy and lead to a degradation of
the effective center of mass energy for $e^+e^-$ collisions, on the
other hand, the lower energy $e^+e^-$ and $\gamma$'s contribute to
background processes.

In general, the full treatment of beam-beam interactions and
beamstrahlung is a complicated many-body problem, and dedicated codes
for the calculation of beam-beam effects exist~\cite{ABEL} or are being
developed~\cite{CAIN}.  However, from the high energy physics
perspective the availability of a simplified analytical or numerical
treatment is desired for the study of interesting processes, (e.g.,
$e^+e^- \to W^+W^-$), at the level of Monte Carlo event generators.

As long as the average number of beamstrahlung photons per beam particle
is much less than unity, the radiation spectrum is given by the
Sokolov-Ternov spectrum \cite{ST86}.  In the case of the proposed linear
colliders, the average number of radiated photons per beam particle at
design luminosity is typically ${\cal O}(1)$.  Therefore, the effects of
multiple radiation of beamstrahlung photons have to be taken into
account.

Multiphoton beamstrahlung has already been discussed in
\cite{BCK91,Chen92} (and references quoted therein).  In this note we
will follow the approach by Chen et al., however, without using any
approximation to the radiation spectrum.  We shall derive a recursive
solution of the evolution equation, which is well suited for a numerical
implementation.

The Monte Carlo algorithm for beamstrahlung presented below is
considered complementary to a full treatment of all beam-beam effects as
performed in \cite{ABEL}.  The main purpose of this Monte Carlo approach
is to provide a simple package that can be incorporated into or
interfaced with Monte Carlo event generators for physics at future
$e^+e^-$ colliders.  In section 2 we will review the electron energy
spectrum under multiphoton beamstrahlung.  Section 3 outlines our Monte
Carlo approach, while in section 4 we present preliminary results.
Further details of the calculation will be presented
elsewhere~\cite{Anl96}.

%%%%%%%%%%%%%%%%%%%%%%%%%%%%%%%%%%%%%%%%%%%%%%%%%%%%%%%%%%%%%%%%%%%%%%%%

\section{Electron Energy Spectrum}

In this section we recapitulate the derivation of the electron energy
spectrum due to multiphoton beamstrahlung in the spirit of Chen et
al.~\cite{BCK91,Chen92}.  We use $\hbar=c=1$.

Let $\psi(x,t)$ be the distribution function of the electron with energy
fraction $x = E/E_0$ at time $t$, normalized such that
\begin{equation}
\label{eq:norm}
\int_0^1 dx \; \psi(x,t) = 1 \; .
\end{equation}
We shall assume that the emission of the photons takes place in an
infinitesimally short time interval.  Therefore the interference between
successive emissions may be neglected, and one can describe the time
evolution of the electron distribution by the rate equation
\begin{equation}
\label{eq:evolution}
\frac{\partial\psi}{\partial t} =
 - \nu(x) \psi(x,t) + \int_x^1 dx' \; F(x,x') \psi(x',t) \; ,
\end{equation}
where the first and second term on the r.h.s.\ correspond to the sink
and source for the evolution of $\psi$, respectively.  $F$ is the
spectral function of radiation, and
\begin{equation}
\nu(x) = \int_0^x dx'' \; F(x'',x)
\end{equation}
represents the average number of photons radiated per unit time by an
electron with energy fraction $x$.  Electrons and positrons from pair
production in photon-photon processes are not taken into account.

The spectral function of radiation is conventionally characterized by
the beamstrahlung parameter $\Upsilon$, which is defined as
\begin{equation}
\label{eq:Upsilon}
\Upsilon = \frac{E_0}{m_e} \frac{B}{B_c} \; ,
\end{equation}
where $B$ is the effective field strength in the beam, and $B_c =
m_e^2/e$ is the Schwinger critical field.  The $e^+e^-$ beams generally
have Gaussian charge distributions, and thus the local field strength
varies inside the beam volume.  However, it has been shown \cite{Chen88}
through integration over the impact parameter and the longitudinal
variations, that the overall beamstrahlung effect can be simply
described as if all particles experience a uniform mean field $B_{mean}$
during an effective collision time $\tau = l_{eff}/2 = \sqrt{3}
\sigma_z$.  In what follows, we will assume the obtained mean
beamstrahlung parameter $\Upsilon_{mean}$ and effective beam length
$l_{eff}$ but drop the subscripts.

The spectral function of synchrotron radiation was derived by Sokolov
and Ternov \cite{ST86} and reads
\begin{eqnarray}
F(x,x') & = &
\frac{\nu_{cl}}{\Upsilon} \frac{2}{5\pi} \frac{1}{x'^2}
\left\{ \int_\eta^\infty du \; K_{5/3}(u) +
\frac{(\xi\eta)^2}{1+\xi\eta} K_{2/3}(u) \right\} \theta(x'-x) \; ,
\end{eqnarray}
where $\xi = (3\Upsilon/2) x'$, $\eta=(2/3\Upsilon) (1/x-1/x')$, and
$K_\nu$ denotes the modified Bessel function of order $\nu$.

We define
\begin{equation}
\nu(x) = \int_0^x dx' \; F(x',x) \equiv \nu_{cl} \cdot U_0(x\Upsilon)
 \; ,
\end{equation}
where $\nu_{cl}$ is the number of photons per unit time in the classical
limit,
\begin{equation}
\nu_{cl} = \frac{5}{2\sqrt{3}} \frac{\alpha m_e^2}{E_0} \Upsilon \; ,
\end{equation}
which is independent of the particle energy for a given field strength
due to (\ref{eq:Upsilon}).

The function $U_0(y)$ behaves as
\begin{equation}
U_0(y) = \left\{
\begin{array}{ll}
1 & y \to 0 \\
const. \times y^{-1/3} \quad & y \to \infty
\end{array} \right. \; ,
\end{equation}
exhibiting the asymptotic approach to the classical limit and the
suppression of radiation in the deep quantum regime, respectively.

Finally, the differential $e^+e^-$ luminosity can be expressed as the
convolution of the effective electron energy distributions of the
colliding beams~\cite{Chen92},
\begin{equation}
\label{eq:diff-lumi}
 \frac{ d{\cal L}(s) }{ds} =
 {\cal L}_0 \int dx_1 dx_2 \; \delta(s-x_1 x_2 s_0) \psi(x_1) \psi(x_2)
 \; ,
\end{equation}
where ${\cal L}_0$ is the nominal luminosity of the collider, including
the enhancement factor due to beam disruption, and the effective energy
distribution $\psi(x)$ is obtained by averaging over the longitudinal
position within the beam,
\begin{equation}
\label{eq:eff-energy}
  \psi(x) = \frac{2}{l} \int_0^{l/2} \psi(x,t) \; .
\end{equation}

%%%%%%%%%%%%%%%%%%%%%%%%%%%%%%%%%%%%%%%%%%%%%%%%%%%%%%%%%%%%%%%%%%%%%%%%

\section{Multiphoton Beamstrahlung by Monte Carlo}

The evolution equation (\ref{eq:evolution}) may be solved by a Monte
Carlo algorithm.  Here we will outline only the basic idea, deferring
the full presentation and discussion to a forthcoming
publication~\cite{Anl96}.

To determine the electron distribution $\psi(x,t)$, let us decompose it
into a suitable set of time independent functions $\phi_n(x)$ using the
ansatz
\begin{equation}
\label{eq:ansatz}
 \psi(x,t) = \sum_{n=0}^\infty
 \phi_n(x) \cdot \frac{(\nu t)^n}{n!} e^{-\nu t} \; ,
\end{equation}
where $\nu$ is a parameter that will be determined below.
The normalization condition (\ref{eq:norm}) translates into
\begin{equation}
\label{eq:norm-phi}
\int_0^1 dx \; \phi_n(x) = 1 \; .
\end{equation}

Plugging the ansatz (\ref{eq:ansatz}) into the evolution equation
(\ref{eq:evolution}) we obtain a recursion relation for the $\phi_n$,
\begin{equation}
\label{eq:recursion}
\phi_n(x) = \int_0^1 dx'
 \left[ \left(1 - \frac{\nu(x)}{\nu}\right) \delta(x-x') +
        \frac{F(x,x')}{\nu} \right] \phi_{n-1}(x')
 \quad \mbox{ for } n \geq 1 \; ,
\end{equation}
with initial condition
\begin{equation}
\phi_0(x) = \psi(x,0) \; .
\end{equation}
If we neglect the intrinsic (Gaussian) energy spread of the accelerator,
the initial condition of the electron distribution is simply given by
\begin{equation}
\label{eq:initial-cond}
\psi(x,0) = \delta(1-x) \; ,
\end{equation}
and thus
\begin{equation}
\phi_0(x) = \delta(1-x) \; .
\end{equation}

The functions $\phi_n(x), n > 0$, can be determined, e.g., by numerical
integration.  Here we suggest a different way to exploit the recursion
relations.  First note that the kernel given by the square brackets in
(\ref{eq:recursion}) is positive for $x' > x$.  Furthermore, since
$\nu(0) \geq \nu(x) \geq \nu(1)$, the kernel is positive if we choose
$\nu = \nu(0) = \nu_{cl}$.  The functions $\phi_n$ will then be
positive, and together with the normalization condition
(\ref{eq:norm-phi}) they will allow for a probabilistic interpretation.

Using our ansatz (\ref{eq:ansatz}), the effective electron energy
distribution (\ref{eq:eff-energy}) reads
\begin{equation}
\label{eq:psi}
\psi(x) = \sum_n \left[ 1 - \frac{\gamma(1+n; N_{cl})}{n!} \right]
\phi_n(x) \; ,
\end{equation}
where $\gamma(1+n; N_{cl})$ is the incomplete gamma function, and
$N_{cl} = \nu_{cl} \cdot l/2$ is the average number of photons per beam
particle radiated during the entire collision, in the classical limit.
Due to the positivity of the term in square brackets in (\ref{eq:psi})
and the functions $\phi_n$, the electron energy distribution can be
determined straightforwardly by a multichannel Monte Carlo
method~\cite{Anl96}.

%%%%%%%%%%%%%%%%%%%%%%%%%%%%%%%%%%%%%%%%%%%%%%%%%%%%%%%%%%%%%%%%%%%%%%%%

\section{Results}
\label{sec:results}

Of the abundance of available parameter sets for future linear $e^+e^-$
colliders, we have taken a hopefully representative selection
from~\cite{Weise} that comprises many of the proposed linear collider
designs.  Table~\ref{tab:NLC-params} lists some of the parameters that
are relevant for our numerical results.  $\Upsilon_{mean}$ was
determined similarly to \cite{Chen92}.

\begin{table*}
\begin{center}
\begin{tabular}{|l|c|c|c|c|c|c|c|}
\hline
Parameter     & SBLC  & TESLA &  NLC  & JLC-S & JLC-X & CLIC  & VLEPP \\
\hline
Luminosity [$10^{33} {\rm cm}^{-2} {\rm s}^{-1}$] &
		5.3   &  6.1  &  6.9  &  4.6  &  5.1  &   ?   &  9.7  \\
Particles/bunch $[10^{10}]$ &
		1.1   &  3.63 &  0.65 &  1.44 &  0.63 &  0.8  &  20   \\
$\sigma_x$ [nm] &
		335   &  845  &  320  &  260  &  260  &  250  &  2000 \\
$\sigma_y$ [nm] &
		15.1  &  18.9 &  3.2  &  3.04 &  3.04 &  7.5  &  3.9  \\
$\sigma_z$ [$\mu$m] &
		300   &  700  &  100  &  120  &   90  &  200  &  750  \\
$\Upsilon_{mean}$ &
		0.049 & 0.029 & 0.090 & 0.213 & 0.120 & 0.072 & 0.064 \\
$N_{cl} = \nu_{cl} \cdot l/2$ &
		1.42  & 1.93  & 0.87  & 2.47  & 1.04  &  1.4  &  4.6  \\
${\cal L}(s>0.99\,s_0)/{\cal L}_0$ &
		0.58  & 0.55  & 0.68  & 0.32  & 0.62  & 0.56  &  0.18 \\
\hline
\end{tabular}
\caption{{\em Selected parameters of some proposed linear collider designs.}}
\label{tab:NLC-params}
\end{center}
\end{table*}

\begin{figure*}[hbt]
\begin{center}
 \mbox{\epsfxsize=14cm\epsffile{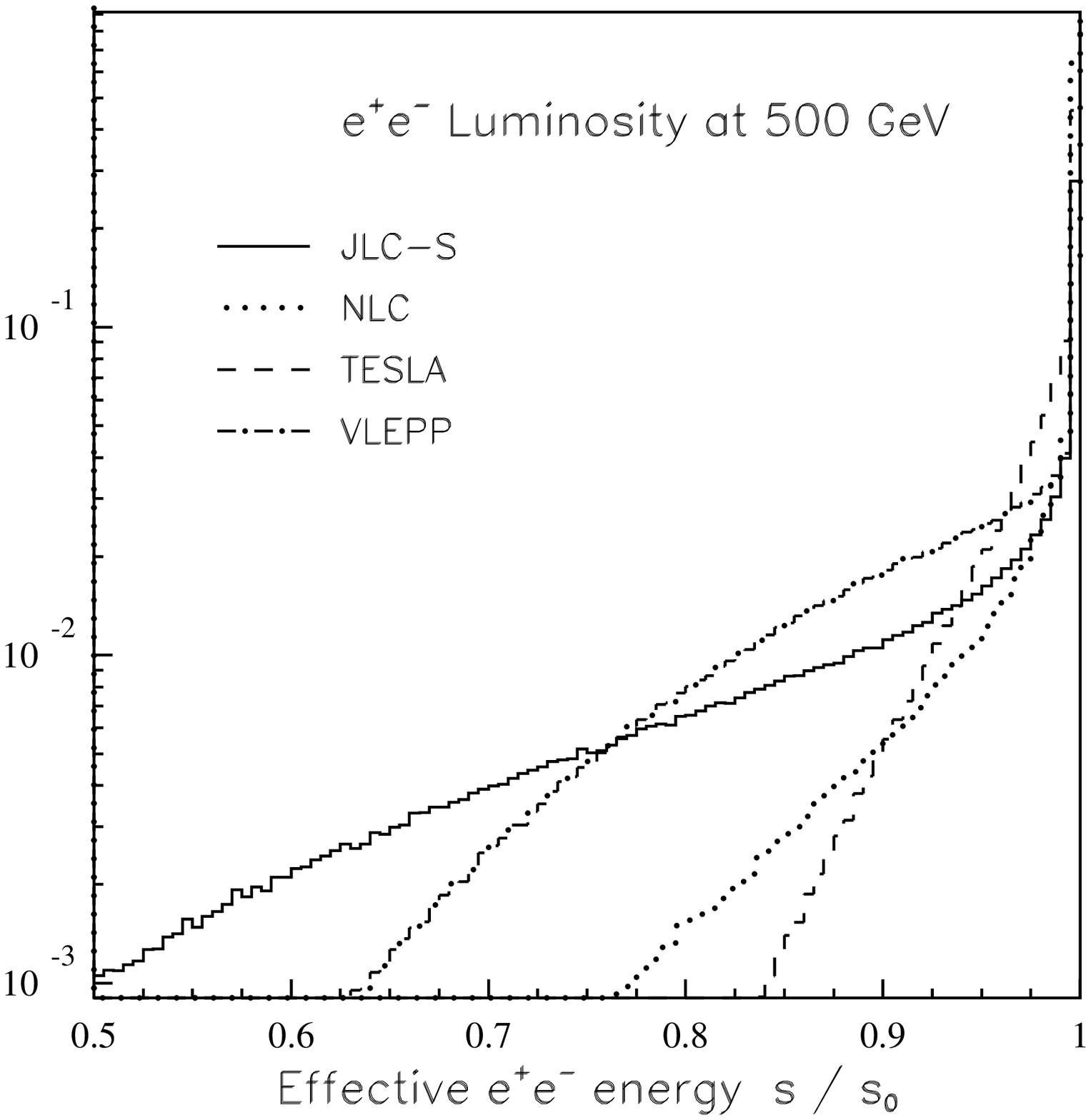}}
\caption{{\em Differential $e^+e^-$ luminosity.}}
\label{fig:luminosity}
\end{center}
\end{figure*}

We have calculated the differential luminosity of the $e^+e^-$ beams
under beamstrahlung, eq.~(\ref{eq:diff-lumi}), using our Monte Carlo
algorithm based on (\ref{eq:recursion}).  The result of the simulation
for the normalized luminosity is shown in figure~\ref{fig:luminosity}.
For the sake of brevity we have selected only a subset of four designs
which essentially cover the interesting parameter range: the TESLA
design with rather large bunch length but small effective beamstrahlung
parameter and a correspondingly rather soft beamstrahlung spectrum, the
JLC-S design with shorter bunches but large $\Upsilon$ and large mean
energy loss due to a hard beamstrahlung spectrum, NLC with short bunches
and moderate $\Upsilon$, having a small number of radiated photons
$N_{cl}$, and VLEPP with a large multiplicity of beamstrahlung photons.
We have also calculated the relative luminosity for collisions with $s >
0.99\, s_0$, the result being displayed in the last line of
table~\ref{tab:NLC-params}.

In high energy $e^+e^-$ processes, besides beamstrahlung one also has to
take into account the radiative corrections to the process under
consideration.  The single most important universal contribution of the
radiative corrections are the leading logarithms $\alpha^n
\log^n(s/m_e^2)$, which are due to QED initial state radiation (ISR).
These large logarithms can be resummed to all orders in the structure
function approach, with the electron structure function \cite{StrFun}:
\begin{equation}
  D(z; \mu^2) =
  \frac{\beta}{2} (1-z)^{\frac{\beta}{2}-1}
  \left(1+\frac{3}{8}\beta \right) - \frac{1}{4}\beta(1+z)
  + {\cal O}(\beta^2) \; ,
\end{equation}
where
\begin{equation}
  \beta = \frac{2\alpha}{\pi} \left( \log\frac{\mu^2}{m_e^2} - 1 \right)
  \; ,
\end{equation}
$\mu^2$ is the factorization scale of order $s$, and $z$ is the electron
energy fraction after initial state radiation.

\begin{figure*}[tb]
\begin{center}
 \mbox{\epsfxsize=14cm\epsffile{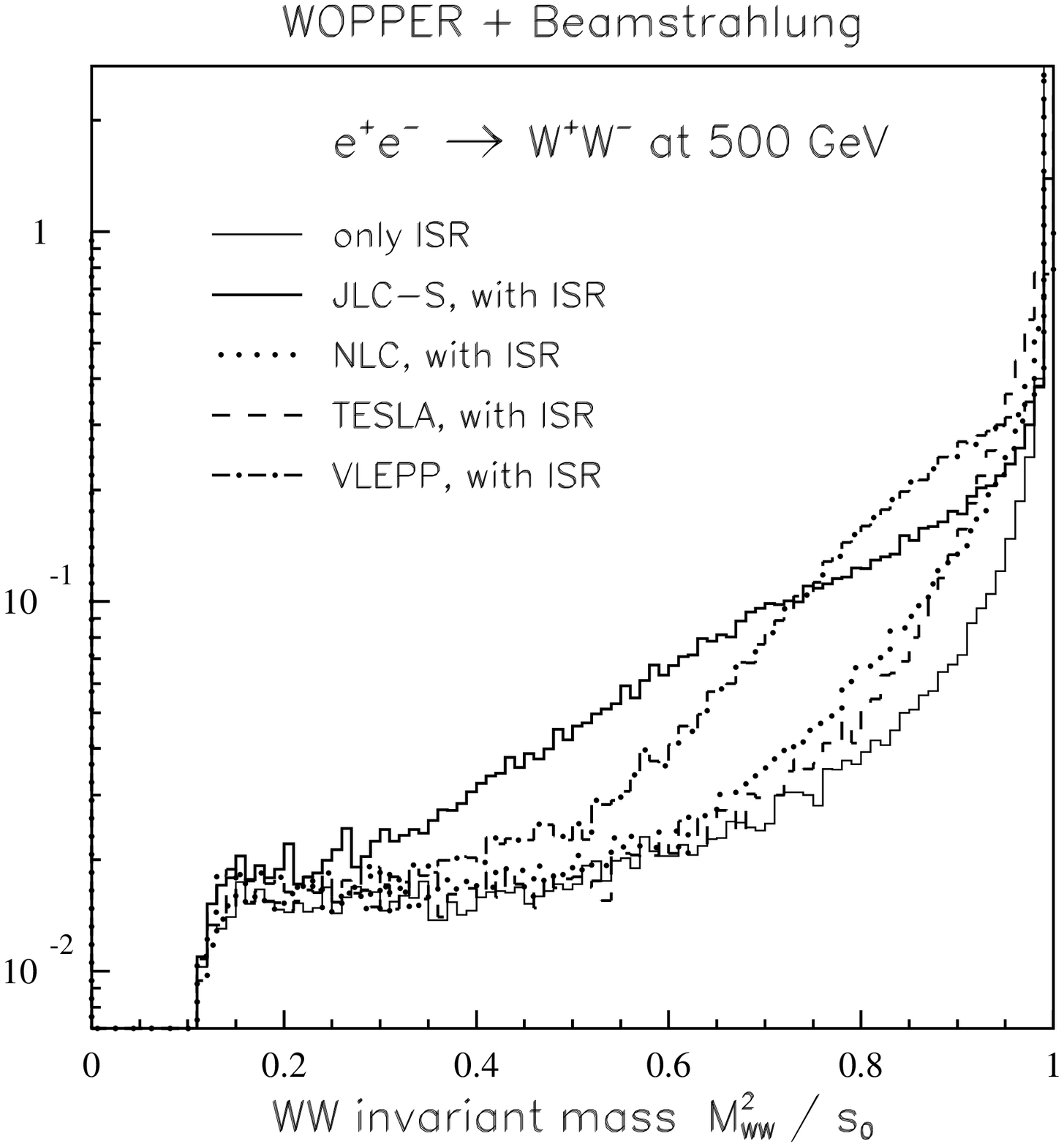}}
\caption{{\em Final state invariant mass distribution with and without
beamstrahlung.}}
\label{fig:WWinvmass}
\end{center}
\end{figure*}

As an application, we have interfaced our Monte Carlo routines for
beamstrahlung with the Monte Carlo event generator {\tt WOPPER}
\cite{WOPPER}.  {\tt WOPPER} simulates the process $e^+e^- \to (W^+W^-)
\to 4f$ including QED radiative corrections in the abovementioned
leading logarithmic approximation with explicit photons.  In
figure~\ref{fig:WWinvmass} we show the normalized invariant mass
distribution $M^2_{WW}$ of the final state (assuming a 100\%
reconstruction efficiency for simplicity),
\begin{equation}
\frac{1}{\sigma_0} \frac{d\sigma}{dM^2_{WW}} =
\int dx_1 dx_2 dz_1 dz_2 \;
 \psi(x_1) \psi(x_2) \; D(z_1;\mu^2)  D(z_2;\mu^2) \cdot
\frac{1}{\sigma_0} \sigma_0(x_1 z_1 x_2 z_2 s) \; ,
\end{equation}
using the same collider designs as in figure~\ref{fig:luminosity}.  The
thin continuous line represents the corresponding distribution from ISR
alone.  For events with a significantly reduced invariant mass the
corrections due to beamstrahlung are numerically much larger than due to
pure ISR, and the shape of the distribution is essentially dominated by
the beamstrahlung spectrum.  This result clearly shows the importance of
beamstrahlung and the need to include beamstrahlung into Monte Carlo
event generators for physics at future $e^+e^-$ colliders.

%%%%%%%%%%%%%%%%%%%%%%%%%%%%%%%%%%%%%%%%%%%%%%%%%%%%%%%%%%%%%%%%%%%%%%%%

\end{document}